\documentclass[sigconf]{acmart}

\usepackage{array} 
\usepackage{multirow} 
\usepackage{rotating} 
\usepackage{enumitem}
\usepackage{amsmath}
\usepackage{caption}
\usepackage{footnote}

\AtBeginDocument{%
  }

\setcopyright{acmlicensed}

\copyrightyear{2024} 
\acmYear{2024} 
\setcopyright{rightsretained} 
\acmConference[ICAIF '24]{5th ACM International Conference on AI in Finance}{November 14--17, 2024}{Brooklyn, NY, USA}
\acmBooktitle{5th ACM International Conference on AI in Finance (ICAIF '24), November 14--17, 2024, Brooklyn, NY, USA}
\acmDOI{10.1145/3677052.3698662}
\acmISBN{979-8-4007-1081-0/24/11}

\begin{document}

\title[Stock Recommendations for Individual Investors]{Stock Recommendations for Individual Investors: A Temporal Graph Network Approach with Mean-Variance Efficient Sampling}


\author{Youngbin Lee}
\orcid{0009-0000-5021-5091}
\email{young@unist.ac.kr}
\affiliation{%
    \institution{Ulsan National Institute of Science and Technology (UNIST)}
    \country{Republic of Korea}
}
\authornotemark[1]

\author{Yejin Kim}
\orcid{0009-0005-6141-9360}
\email{kimyejin99@unist.ac.kr}
\affiliation{%
    \institution{Ulsan National Institute of Science and Technology (UNIST)}
    \country{Republic of Korea}
}
\authornotemark[1]


\author{Javier Sanz-Cruzado}
\orcid{0000-0002-7829-5174}
\email{javier.sanz-cruzadopuig@glasgow.ac.uk}
\affiliation{
    \institution{University of Glasgow}
    \country{United Kingdom}
}

\author{Richard McCreadie}
\orcid{0000-0002-2751-2087}
\email{richard.mccreadie@glasgow.ac.uk}
\affiliation{
    \institution{University of Glasgow}
    \country{United Kingdom}
}
\authornotemark[2]

\author{Yongjae Lee}
\orcid{0000-0002-5411-4340}
\email{yongjaelee@unist.ac.kr}
\affiliation{%
    \institution{Ulsan National Institute of Science and Technology (UNIST)}
    \country{Republic of Korea}
}
\authornotemark[2]




\thanks{\textsuperscript{*}These authors contributed equally.}
\thanks{\textsuperscript{$\dagger$}Co-corresponding authors.}

\definecolor{hotpink}{HTML}{FF69B4}
\begin{abstract}
Recommender systems can be helpful for individuals to make well-informed decisions in complex financial markets. While many studies have focused on predicting stock prices, even advanced models fall short of accurately forecasting them. Additionally, previous studies indicate that individual investors often disregard established investment theories, favoring their personal preferences instead. This presents a challenge for stock recommendation systems, which must not only provide strong investment performance but also respect these individual preferences. To create effective stock recommender systems, three critical elements must be incorporated: 1) individual preferences, 2) portfolio diversification, and 3) the temporal dynamics of the first two. In response, we propose a new model, Portfolio Temporal Graph Network Recommender
\texttt{PfoTGNRec}, which can handle time-varying collaborative signals and incorporates diversification-enhancing sampling. On real-world individual trading data, our approach demonstrates superior performance compared to state-of-the-art baselines, including cutting-edge dynamic embedding models and existing stock recommendation models. Indeed, we show that \texttt{PfoTGNRec} is an effective solution that can balance customer preferences with the need to suggest portfolios with high Return-on-Investment.
The source code and data are available at \textcolor{hotpink}{\url{https://github.com/youngandbin/PfoTGNRec}}.

\end{abstract}




\maketitle

\section{Introduction}
In recent years, there has been a significant increase in the number of individual investors participating in the stock market. According to \cite{chang2023changes}, about 58\% of U.S. households owned stocks in 2022, up from 53\% in 2019, marking the highest growth trend in recent history. This surge in participation highlights the growing interest in stock market investment among individual investors.

Despite this increasing engagement, individual investors often exhibit irrational investment behaviors that negatively impact their returns. Common behaviors include overconfidence, the disposition effect, lottery preference, and herding \cite{ngoc2014behavior}. These tendencies result in investment returns that are generally lower than the market average, with the average investor significantly underperforming the S\&P 500 over time \cite{dalbar_qaib}.


There are many established methods for enhancing portfolio performance, one of the most notable being Modern Portfolio Theory (MPT) \cite{markowitz1952jf}. MPT posits that an investor can achieve higher returns for a given level of risk, or reduce risk for a given level of expected return, by selecting a mix of assets. This is accomplished via the diversification effect, which combines assets with low or negative correlations. Such diversification is effective in reducing overall portfolio risk, and it can be further enhanced by using machine learning techniques \cite{lee2023overview}. While MPT has been the foundation of investment management of most institutions \cite{kim2021mean}, individual investors typically do not follow these sophisticated methods \cite{kim2020personalized, kim2022goal}. Instead, their investment decisions are often driven by personal preferences, which are influenced by various factors such as psychological biases, news, and peers.

\begin{figure*}[!htp] 
    \centering
    \includegraphics[width=\textwidth, height=0.8\textheight, keepaspectratio]{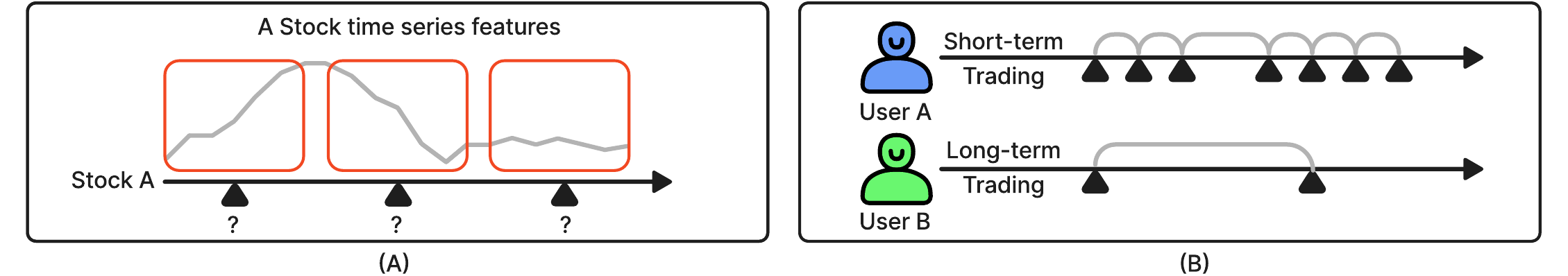}
    \caption{The importance of temporal aspects in stock recommender systems. (A) Various features of stocks would be different based on the timing of recommendations. (B) Contrasting behaviors between user A, who engages in short-term trading, and user B, who holds stocks for a long period.}
    \label{fig:introduction_example}
\end{figure*}

Given these varied influences and the tendency for irrational investment behaviors, there is a clear need for a stock recommendation system. Such a system can guide individual investors, helping them make more disciplined and informed investment decisions. Indeed, in theory, by leveraging advanced recommendation models, it should be possible to capture both user preferences and market dynamics more effectively and concurrently, ultimately improving the investment outcomes for individual investors.


Creating an effective stock recommendation system involves several key considerations.

The first is \textbf{individual preference}.
In essence, individual investment behaviors are highly heterogeneous \cite{khan2017impact, HWANG2024105481, hwang2023identifying}. 
Individual investors often navigate their unique paths, like interpreting and assessing information obtained from media and peers, choosing a few stocks instead of adhering to well-diversified portfolios \cite{kim2021recent}.
For instance, a study analyzing Robinhood investors \cite{welch2022wisdom} revealed the phenomenon of "experience holding," where investors find pleasure in simply holding certain stocks that are chosen based on not purely cash-flow-based perspectives. \citet{bhattacharya2012unbiased} found that most retail investors do not follow unbiased financial advice from experts. They quoted the famous saying, \textit{"You can lead a horse to water, but you can't make it drink."} That is, even though we can build a model that exhibits better investment performance, most retail investors would not take it if they do not like it. 

However, most existing studies on investment recommendation only consider aspects related to the prices of financial assets \cite{paranjape2013stock,nair2015stock,tu2016investment}. 
There are two problems with price-based recommendations. First, it is almost impossible to provide accurate predictions of financial asset prices. Even the most sophisticated models exhibit accuracy around 52 to 57\%, which is not enough to generate positive returns after transaction fees \cite{yoo2021accurate}. Second, they do not consider individual preferences. As noted before, many individuals are unlikely to follow recommendations that do not align with their tastes.

\looseness -1 The second is \textbf{investment performances}, specifically, the diversification effect. No matter how well a model aligns with individual preferences, it is of no use if investment performance is poor. According to the modern portfolio theory originated from \cite{markowitz1952jf}, diversification involves including stocks with low correlations in a portfolio to reduce risk and achieve stable returns. The diversification is crucial in investment management because the price prediction of financial assets would naturally include substantial error, and it has been the key success factor of most institutional investors \cite{kim2021mean}.

However, the tricky point in stock recommendation is that the first two key aspects, individual preference, and investment performance, have a trade-off relationship. In experiments on 12 financial asset recommender (FAR) systems \cite{sanz2022transaction}, it was concluded that transaction-based and profitability-based metrics are not interchangeable. FAR systems that learned from past pricing history showed high performance in return but performed poorly in individual preference, i.e. near zero. Conversely, FAR systems that learned from past transactions demonstrated good performance in individual preference but showed a downward trend in return.
This shows that 'customers are not always right' in stock recommendations. Therefore, \textit{it is inevitable that a trade-off between preference and profitability will need to be made if we are to achieve better investment performance for stock recommendation}.

Lastly, the \textbf{temporal nature} of stock features and user preferences is important. 
Figure \ref{fig:introduction_example} illustrates why the temporal aspect should be considered in stock recommendation.
Figure \ref{fig:introduction_example} (A) shows that even the same stock can have very different characteristics depending on the timing of recommendations. If the recommendation is happening at a time point around the first red box, Stock A would seem like a good option. However, it would be better not to recommend Stock A during the second and third red boxes.
In Figure \ref{fig:introduction_example}  (B), there are two contrasting investment behaviors: user A engages in short-term trading, while user B holds stocks for an extended period. Thus, it is essential to consider the temporal dynamics of user behaviors.

In this paper, we propose a stock recommender system called \textbf{P}ort\textbf{fo}lio \textbf{T}emporal \textbf{G}raph \textbf{N}etwork \textbf{Rec}ommender (\texttt{PfoTGNRec}). The proposed model is based on a temporal graph network, developed by \cite{rossi2020temporal}, to extract time-varying collaborative signals (key aspects 1 and 3: individual preference and temporal nature). Further, we incorporate MVECF \cite{chung2023mean} method in sampling contrastive pairs to enhance the diversification effect (key aspect 2: investment performance). Through experiments, we demonstrate that our model is the most effective in improving investment performance while capturing user preferences, achieving a 3.45\% improvement in a comprehensive combined metric compared to the best model among various baselines, including recently developed dynamic graph embedding models and existing stock recommendation models.

\section{Related Works} 

\subsection{Stock Recommendations} Collaborative filtering (CF), which leverages historical user-item interactions, is one of the fundamental and most successful techniques in recommender systems. Methodologies include matrix factorization \cite{koren2009matrix} that decomposes user-item interaction matrix to capture latent relationships between users and items, and Bayesian personalized ranking (BPR) \cite{rendle2012bpr} that operates by determining personalized ranking of items based on user preferences. 

\looseness -1 For stock recommendations, \cite{swezey2018large} propose a CF-based model that takes into account both individual preferences and portfolio diversification. However, CF and portfolio optimization are performed in distinct steps, and such a heuristic approach would lead to sub-optimal results.
\cite{chung2023mean} were the first to develop a holistic model that can effectively handle the trade-off between individual preferences and investment performance. They incorporated modern portfolio theory into a matrix factorization model, as well as developed an associated ranking loss function which can be applied to more advanced models (e.g., GNN-based models). However, all these models do not consider the temporal dynamics of stock features and user preferences.

\looseness -1 On the other hand, there have been several attempts to adapt temporal models for stock recommender systems~\cite{feng2019temporal,gao2021graph,wang2022mg}. However, they focus predominantly on price prediction without consideration of the individual preferences of users.
Meanwhile, \cite{ghiye2023adaptive,takayanagi2023personalized} address the dynamic characteristics of financial markets and user preference, but they fail to systemically address the diversification effect. In contrast, our model aims to simultaneously consider both user preferences and the diversification effect, optimizing stock recommendations holistically.

\subsection{Dynamic Graph Learning} 
In recommender systems, many methodologies utilize graph convolutional networks (GCNs) \cite{kipf2016semi}. This is because the user-item interactions form a graph structure, allowing effective representation learning from such graphs. For example, NGCF \cite{wang2019neural} leverages collaborative signals in high-order connectivities. LightGCN \cite{he2020lightgcn} is specifically designed to enhance scalability, resulting in accelerated training and inference times.

Unlike the typical graph neural networks (GNNs) that learn node embeddings in static graphs, learning embeddings in dynamic graphs where connections change over time requires considering the temporal aspect. For instance, TGAT \cite{xu2020inductive} introduces a time encoding technique upon GAT \cite{velivckovic2017graph}, which is a graph attention mechanism applied to static graphs. In addition, TGN \cite{rossi2020temporal} proposes a more general framework that can incorporate node-wise temporal features. This is an encoder that generates node embeddings at each time step. 
While this framework has been utilized in various graph tasks, there has been no research applying it to recommender systems thus far. In this study, we aim to leverage this framework for recommendation purposes. Comparatively, there have been few works that attempt \emph{dynamic} graph embedding in recommender systems. For example, TGSRec \cite{fan2021continuous} introduces a temporal collaborative Transformer to explicitly model the temporal effects of interactions. Meanwhile, DGEL \cite{tang2023dynamic} refines embeddings based on previous time related embeddings. However, they rely on time encoding without the inclusion of an explicit memory updater, limiting their ability to effectively capture the node history. In contrast, our model utilizes the TGN framework for recommendation task to leverage its capabilities for explicitly embedding node memories with its strong embedding performance.


\section{Preliminaries} 

\indent\indent \textbf{Problem Definition:}
Let us define the task associated with stock recommendations. We denote the set of users as $U = \{u_1, u_2, ... u_{|U|}\}$, the set of items (stocks) as $V = \{v_1, v_2,...,v_{|V|}\}$, and the set of time points as $T = \{t_1, t_2, ... t_{|T|}\}$. Then, a user-item interaction can be represented as $y_{u,v}^t$. If user $u$ purchases the item $v$ at time $t$, then $y_{u,v}^t=1$; otherwise $y_{u,v}^t=0$. Our primary goal is to predict the value of $y^{t}_{u,v}$. Ultimately, for each user and time, the model aims to recommend the top-k items, leading to a personalized and time-sensitive set of stock recommendations that can improve the portfolio's investment performance.

\textbf{Continuous Time Dynamic Graph:}
We construct a dynamic graph with user-item interactions, changing its structure over time. 
We define our continuous-time bipartite graph as $\mathcal{G}(T) = (\mathcal{V}, \mathcal{E}_T)$. Here, $\mathcal{V}$ represents the set of user and item nodes. 
$\mathcal{E}_T$ denotes the temporal set of edges. Each edge in $\mathcal{E}_T$ is characterized by a tuple $e = (u, v, t, \mathbf{e}_{uv})$, consisting of a user node $u$, an item node $v$, a timestamp $t$, and an edge feature $\mathbf{e}_{uv}$. 
If a user interacts with the same item multiple times, each interaction is represented as a distinct edge in the graph. This approach enables the construction of a dynamic graph that accurately captures the evolving relationships between nodes over time.



\begin{figure*}[htp]  
    \centering
    \includegraphics[width=\textwidth, height=\textheight, keepaspectratio]{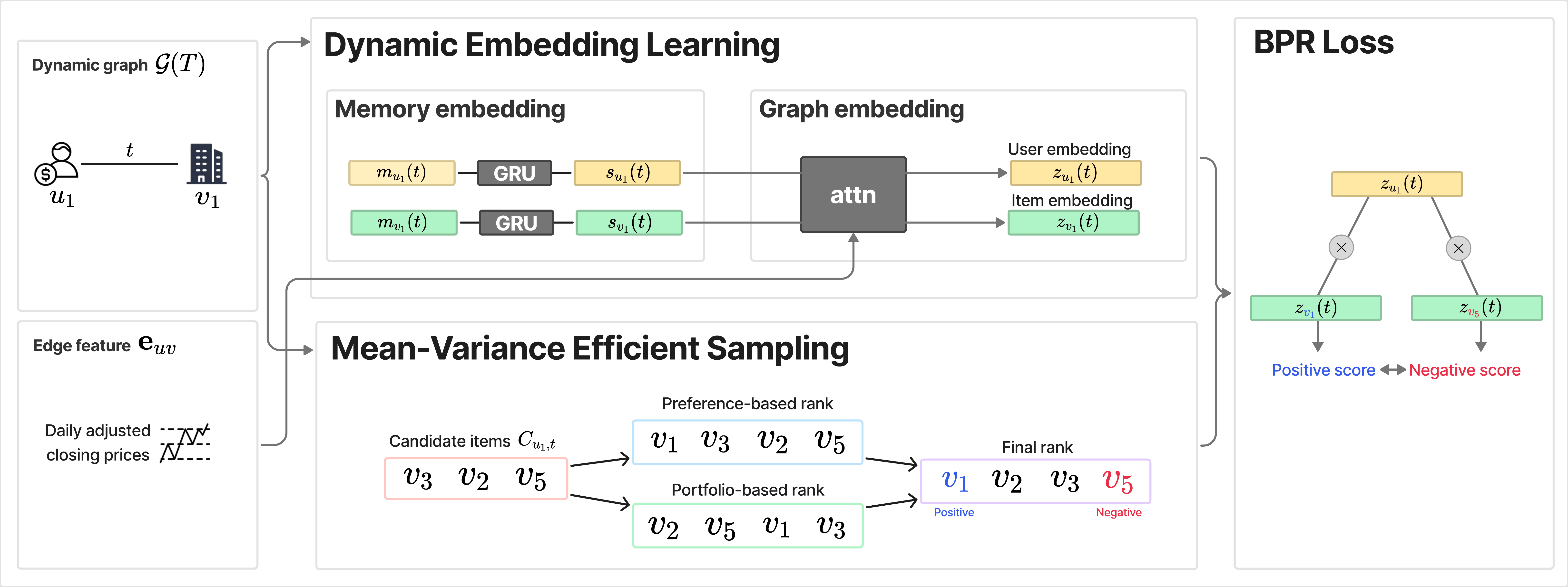}
    \caption{Model architecture of our proposed \texttt{PfoTGNRec}.}
    \label{fig:data}
\end{figure*}

\section{Method} 

We present the \texttt{PfoTGNRec} model, which consists of three integral components: (1) Dynamic embedding learning, where we utilize TGN encoder to effectively learn the evolving characteristics of user-item interactions, (2) Mean-variance efficient sampling, which involves strategic item sampling and designing contrastive pairs to enhance the user portfolio, and (3) Optimization, where the model is trained with Bayesian Personalized Ranking (BPR) loss. 

\subsection{Dynamic Embedding Learning}

First, we learn node embeddings from our dynamic graph constructed from user-item interactions, which are later used when calculating recommendation scores.

\subsubsection{Memory embedding}
We generate memory embeddings for each node to capture the dynamic nature. The process begins with the extraction of information from each node, termed as ``message''. 
In the case of an interaction between source node ($i$) and destination node ($j$) at time $t$,  two messages are computed:
\begin{equation}
    m_{i}(t)  = s_i\left(t^{-}\right) \| s_j\left(t^{-}\right) \| \Delta t \| \mathbf{e}_{i j}
\end{equation}
and 
\begin{equation}
    m_{j}(t)  = s_j\left(t^{-}\right) \| s_i\left(t^{-}\right) \| \Delta t \| \mathbf{e}_{j i}
\end{equation}
Here, $\|$ is a concatenation operator, $s_i\left(t^{-}\right)$ and $s_j\left(t^{-}\right)$ represent the memory at the previous time step for the source and destination nodes, respectively, $\Delta t$ is the time interval $t - t^{-}$, and $\mathbf{e}_{ij}$ is the edge feature.

In the memory update process, a recurrent neural network approach is employed to update the memory of a node following each interaction that involves the node itself. Specifically, GRU \cite{cho2014learning}  is utilized in our model, and the memory is updated as follows:
\begin{equation}
    s_i(t)=GRU(m_i(t), s_i(t^-))
\end{equation}

\subsubsection{Graph embedding}
In this module, temporal embeddings for a dynamic graph are generated. In specific, embeddings are created for each node at time step $t$. Graph attention is utilized to effectively learn the connectivity between nodes. A node embedding can be represented as:
\begin{equation}
    \mathbf{z}_i(t)=\sum_{j \in n_i^k(t)} attn\left(s_i(t), s_j(t), \mathbf{e}_{i j}\right) 
\end{equation}

where $attn$ refers to the graph attention mechanism as described in \cite{rossi2020temporal}, $s_i(t)$ and $s_j(t)$ represent the memory, 
$\mathbf{e}_{ij}$ is the edge feature, and the neighborhood set of node $i$, denoted as $n_i^k(t)$, refers to the $k$-hop temporal neighbors connected at time $t$.

\subsection{Mean-Variance Efficient Sampling}

Unlike conventional recommender systems that sample contrastive pairs based on user purchase history or user-item similarity, we take into account the portfolio diversification effect. In other words, while performing positive and negative sampling based on user-item interactions, we are motivated by MVECF \cite{chung2023mean} to reflect rankings according to mean-variance when sampling items.
Consider user $u$ bought item $v$ at time $t$. At this interaction point, a user's current portfolio $PO_{u,t}$ consists of items that the user holds at $t$, representing a collection of various stocks. Then, we randomly sample a set of candidate items $C_{u,t}$ that do not belong to the user portfolio $PO_{u,t}$, from the item set $V$.
\begin{equation}
C_{u,t} \leftarrow \text{sample}(V - PO_{u,t})
\end{equation}

Now, we create two ranked lists using the candidate items: (1) a preference-based list and (2) a portfolio-based list. In (1) preference-based list, the item that the user has actually purchased at that time is ranked first, and candidate items are ranked randomly in the remaining positions. (2) The portfolio-based list ranks the items based on their profitability and volatility, regardless of the user's preferences. This is done by calculating the mean-variance score for each item and ranking them in descending order of their scores.


The mean-variance score is designed to consider the effectiveness of adding an item to the existing portfolio in enhancing diversification effects. The modified target rating in the MVECF, $y^{MV}_{ui}$, is calculated as follows:

\begin{equation}
\begin{aligned}
y^{MV}_{ui} & =\frac{\frac{\mu_i}{\gamma}-\frac{1}{2} \sum_{j: j \neq i} \frac{1}{\left|y_u\right|} \sigma_{i j}}{\sigma_i^2}
\end{aligned}
\end{equation}

Here, 
$\gamma$ is a hyperparameter for risk-aversion level and $|y_u|$ represents the number of holdings of user u. 
We calculated the mean return and variance of items, denoted by $\mu$ and $\sigma$ respectively, based on the prices over the next 30 days from the point of calculating the MV score.
As the formula indicates, 
$y^{MV}_{ui}$ assigns higher values to items that increase returns while decreasing risk, when added to the user's current portfolio.

To get the final rank of items, we combine the preference-based list and portfolio-based list by calculating a weighted sum of the rankings from them. Here, the weight $\lambda_{MV}$ ranges from 0 to 1.
Finally, we choose positive and negative items from the final rank. The items ranked at the top are sampled as positive items $P_{u,t}$, while the items ranked at the bottom are sampled as negative items $N_{u,t}$. In this study, we selected one top-ranked item as positive and three bottom-ranked items as negative.
\begin{align}
P_{u,t} = \text{top-ranked items from the final rank} \\
N_{u,t} = \text{bottom-ranked items from the final rank} 
\end{align}

\subsection{BPR Loss}

Following the typical recommender systems, we employ the BPR loss to train the model. At the time when the interaction takes place, we sample positive and negative items with mean-variance efficient sampling. Then, BPR loss is applied to calculate scores for pairs of positive and negative items.


\begin{equation}
    \mathcal{L}_{B P R}=\sum_{(u, p, n, t) \in D}-\log \sigma\left(\mathbf{z}_u(t)^T \mathbf{z}_p(t)-\mathbf{z}_u(t)^T \mathbf{z}_n(t)\right)
\end{equation}

In this equation, \(D\) denotes the edge set, which is derived from \(\mathcal{E}_T\), \(u\) represents user, $p$ represents the positive item selected from \(P_{u,t}\), and $n$ is the negative item selected from \(N_{u,t}\)


\section{Experiment}

In this section, we explain how we assessed the performance of our proposed model using a Greece trading dataset collected from real customer investment transactions. We formulated our experimental questions based on two pivotal aspects that ought to be considered in \texttt{PfoTGNRec}: recommendation and portfolio performance. We aim to answer the following research questions:
\begin{itemize}
    \item \textbf{RQ1:} Can \texttt{PfoTGNRec} provide a better trade-off between recommendation and investment performance than past stock recommendation algorithms?
    \item \textbf{RQ2:} How effective is \texttt{PfoTGNRec} in comparison to past stock recommendation algorithms on predicting individual customer investments (recommendation performance)?
    \item \textbf{RQ3:} How profitable are the recommendations provided by \texttt{PfoTGNRec} in comparison to past stock recommendation algorithms (portfolio or investment performance)?
    \item \textbf{RQ4:} How do \texttt{PfoTGNRec} hyperparameters affect its investment and recommendation performance?
\end{itemize}

\subsection{Experimental Settings}

\subsubsection{Dataset} We conduct experiments using individual investor transaction dataset, provided by National Bank of Greece~\citep{sanz2024dataset}. This dataset includes real transaction data of users and represents a snapshot of the Greek market. The data spans from January 2018 to November 2022 comprising user buy orders during this period. To exclude abnormal transactions, we remove stocks with highly unstable price movements. We use daily adjusted closing prices for the temporal features of stocks, retrieved from an open source Python package \textit{yfinance}.

For the sake of conducting stable experiments, we perform some filtering on items. 
We use stock price data from Yahoo Finance, and stocks and dates that do not exist in Yahoo Finance are excluded from the data.
Additionally, to eliminate stocks that have been halted in trading, we remove stocks with no price changes for 30 consecutive days. 
As a result, the average number of interactions per user is 18.24, with a median of 5. For the number of interactions per item, the average is 1,653.09, with a median of 393.

To obtain real-time user portfolios of users during the data period, we utilize buy and sell orders along with the quantities of stocks ordered.
Portfolios represent users' stock holdings for each interaction as the set of stocks held up to the day before.
The average number of stocks in user portfolios is 6.26, with a median of 5. The minimum number of stocks is 0, and the maximum is 47. Most users hold fewer than 10 stocks in their portfolios.

For the edge features that change over time, we use daily adjusted closing prices of the most recent 30 trading days before each interaction.

For the data split, we utilize a chronological approach based on interaction timestamps to partition the dataset into training, validation, and testing sets. This division follows a ratio of 8:1:1, which preserves the temporal order of interactions. Ultimately, we use data consisting of 8,337 users, 92 stocks, and 152,084 interactions.

\subsubsection{Baseline}  
We have selected the baseline models based on the following three categories.

\textbf{Recommender models}: We compare our model with competitive transaction-based algorithms, both static and dynamic. The \textit{static} methods include Pop, BPR \cite{rendle2012bpr}, WMF \cite{hu2008collaborative}, LightGCN \cite{he2020lightgcn} which is a static graph learning method, and SGL \cite{wu2021self} which leverages a self-supervised learning approach. For \textit{dynamic} methods, we consider state-of-the-art dynamic graph learning models including 
DyRep \cite{trivedi2019DyRep}, Jodie \cite{kumar2018learning}, 
TGAT \cite{xu2020inductive}, and TGN \cite{rossi2020temporal}. While most dynamic methods have not been utilized for recommendation tasks, we adapt their original architectures to recommendation task by incorporating negative sampling during training and applying BPR loss.

\textbf{Price-based models}: We include risk-return approaches that focus solely on prices rather than transactions. Return and Sharpe model refer to non-personalized models recommending stocks that had the best return and Sharpe ratio over the 30 days before the start of the testing period, respectively. Even the most sophisticated stock price forecasting models (e.g., \cite{yoo2021accurate}) show an accuracy around 55\%, these simple models can serve as good proxies of such models.

\textbf{Stock recommendation models}: We consider the two most advanced stock recommendation models, which are the two-step method \cite{swezey2018large} and MVECF \cite{chung2023mean}. Both can be regarded as \textit{static} methods. 

\subsubsection{Evaluation (Recommendation)}
For the evaluation of recommendation performance, we employ the Hit Ratio (HR) and Normalized Discounted Cumulative Gain (NDCG). All models follow an interaction-based ranking strategy, consistent with the settings in \cite{kumar2018learning}. In other words, for each testing interaction $(u, v, t)$, a list of recommended items was generated. For static models that cannot provide different recommendations for each test interaction, the same item set ranked within the train period is used throughout all test periods. 
To evaluate the performance, we utilize items from the entire item set, excluding those that are in the user's portfolio at each time point.

\subsubsection{Evaluation (Investment)}
To evaluate investment performance, we utilize return and Sharpe ratio. 
For all models, we compare the user's original portfolio with the portfolio after the recommendation for each testing interaction $(u, v, t)$. We constructed the recommended portfolio by adding the top $K$ stocks with the highest recommendation scores to the original portfolio.

In specific, we measure the improvement of investment performance in two ways. First, difference.
The difference in the Sharpe ratio and return is denoted as \(\triangle SR = SR - SR_{\text{init}}\) and \( \triangle R = R - R_{\text{init}}\), respectively. Here, terms appended with “init” represent the original portfolio (before recommendation), whereas those without the suffix refer to the recommended portfolio (after recommendation). These values are calculated for all users and then they are averaged.
Second, the percentage of users whose investment performance becomes better after the recommendation. The improvement percentage in the Sharpe ratio and return are expressed as \(P(SR)=P(SR > SR_{\text{init}})\) and \(P(R)=P(R > R_{\text{init}})\).
To evaluate the actual portfolio performance in the stock market when a stock is recommended, we employ out-of-sample assessments. That is, at the testing point, the investment performance is calculated based on the returns over the next 30 days.

\subsubsection{Configuration}

We train all models for 20 epochs and the reported results are based on the test data with the best performing model selected within the validation set.
For model selection, we use NDCG@5 for recommender, price-based models, and a holistic approach for stock recommendation models and our model, considering both recommendation and investment performance. This is achieved by employing the validation data to independently rank the models based on their performance in NDCG@5 and $P(SR)$@5. Then, the averages of these rankings are used to determine the final model. 
For a fair comparison, we conduct hyperparameter tuning for all models. 


\subsection{Combined Recommendation and Portfolio Performance (RQ1)}

\begin{figure}[ht]
    \centering
    \includegraphics[width=0.9\linewidth]{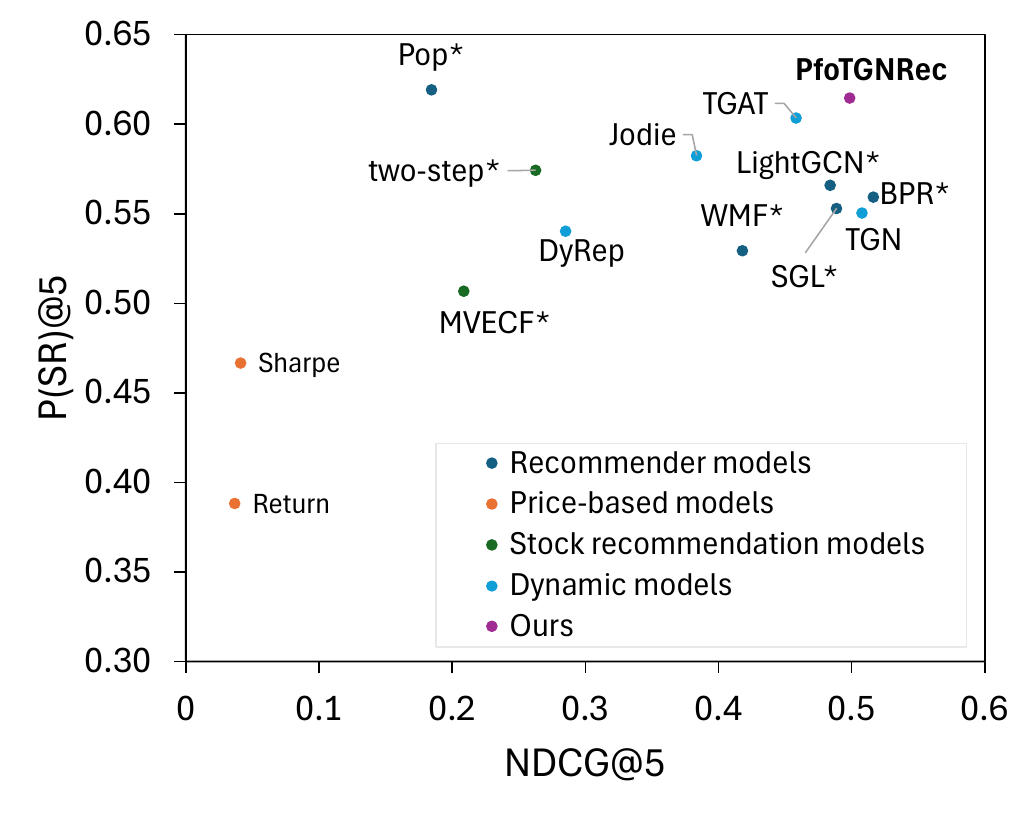}
    \caption{Visualization of Comparison of Both Recommendation and Portfolio Performance}
    \label{fig:model_performance_figure}
\end{figure}

For a comprehensive evaluation of user preferences and portfolio performance, we select two representative metrics: NDCG@5 and P(SR)@5. NDCG@5 is a recommendation performance metric that measures how highly the items actually purchased by the user are ranked on the list of recommended items. P(SR)@5 is an investment performance metric that measures the proportion of users who experienced an improvement in their portfolio Sharpe ratio. These metrics are visualized in Figure \ref{fig:model_performance_figure}. Since higher values for both metrics indicate better performance, models positioned at the outermost points in the graph exhibit the most balanced performance. As shown in Figure \ref{fig:model_performance_figure}, our model demonstrates the best performance.

\begin{table}[ht]
    \centering
    \begin{tabular}{lccccc}
        \hline
        Model & $\alpha=0.3$ & $\alpha=0.4$ & $\alpha=0.5$ & $\alpha=0.6$ & $\alpha=0.7$ \\
        \hline
        Pop & 0.3149 & 0.3584 & 0.4019 & 0.4454 & 0.4889 \\
        WMF & 0.4516 & 0.4627 & 0.4738 & 0.4850 & 0.4961 \\
        BPR & \underline{0.5294} & \underline{0.5337} & \underline{0.5380} & 0.5423 & 0.5466 \\
        LightGCN & 0.5087 & 0.5169 & 0.5250 & 0.5332 & 0.5414 \\
        SGL & 0.5081 & 0.5145 & 0.5210 & 0.5274 & 0.5338 \\
        \hline
        Return & 0.1422 & 0.1774 & 0.2126 & 0.2477 & 0.2828 \\
        Sharpe & 0.1688 & 0.2113 & 0.2539 & 0.2965 & 0.3390 \\
        \hline
        MVECF & 0.2981 & 0.3279 & 0.3578 & 0.3876 & 0.4174 \\
        two-step & 0.3563 & 0.3875 & 0.4186 & 0.4497 & 0.4809 \\
        \hline
        DyRep & 0.3617 & 0.3872 & 0.4128 & 0.4383 & 0.4638 \\
        Jodie & 0.4434 & 0.4632 & 0.4831 & 0.5030 & 0.5228 \\
        TGAT & 0.5021 & 0.5166 & 0.5311 & \underline{0.5456} & \underline{0.5601} \\
        TGN & 0.5207 & 0.5250 & 0.5292 & 0.5335 & 0.5378 \\
        \hline
        \texttt{PfoTGNRec} & \textbf{0.5334} & \textbf{0.5450} & \textbf{0.5566} & \textbf{0.5683} & \textbf{0.5799} \\
        \hline
    \end{tabular}
    {\raggedright \footnotesize \textit{Note}: The best and the second best performing models are highlighted in bold and underline, respectively.\par}
    \vspace{0.5em}
    \caption{Comparison of Models Based on Weighted Metric of Recommendation and Portfolio Performance}
    \label{tab:model_performance}
\end{table}

To provide a precise numerical comparison, we use a combined metric of NDCG@5 and P(SR)@5. We calculate the weighted average of these two metrics using the weight $\alpha$. Specifically, we compute 
\begin{equation}
    m@5(\alpha) = NDCG@5 \times (1-\alpha) + P(SR)@5 \times \alpha
\end{equation}
and vary $\alpha$ from 0.3 to 0.7 to observe performance across different balances. Table \ref{tab:model_performance} displays the overall performance compared to the baseline for various values of $\alpha$. Our model consistently outperforms all baseline models across both recommendation and investment metrics. Therefore, it is evident that our model offers the most balanced approach, enhancing investment performance while reflecting individual preferences.

\subsection{Recommendation Performance (RQ2)}

\begin{table*}
    \centering
    \resizebox{\textwidth}{!}{
    \begin{tabular}{lclccccclcccccccc}
        \cmidrule{2-5} \cmidrule{7-14}
        & \multicolumn{4}{c}{Recommendation effectiveness} & & \multicolumn{8}{c}{Portfolio performance}\\
        \cmidrule{1-5} \cmidrule{7-14}
        Model & HR@3 & HR@5 & NDCG@3 & NDCG@5 & & P(R)@3 & P(R)@5 & P(SR)@3 & P(SR)@5 & $\Delta$R@3 & $\Delta$R@5 & $\Delta$SR@3 & $\Delta$SR@5 \\
        \cmidrule{1-5} \cmidrule{7-14}
        Pop* & 0.1586 & 0.2787 & 0.1355 & 0.1845 & & 0.5174 & \textbf{0.5479} & 0.5670 & \textbf{0.6193} & -0.003 & 0.0106 & 0.1860 & \underline{0.3533} \\
        WMF* & 0.4654 & 0.5588 & 0.3797 & 0.4183 & & 0.4561 & 0.4417 & 0.5228 & 0.5294 & -0.0212 & -0.0379 & 0.0374 & 0.0408 \\
        BPR* & \underline{0.5635} & 0.6538 & \textbf{0.4794} & \textbf{0.5166} & & 0.5234 & 0.4970 & 0.5595 & 0.5594 & 0.0064 & -0.0079 & 0.1499 & 0.1555 \\
        LightGCN* & 0.5378 & 0.6399 & 0.4419 & 0.4841 & & 0.5333 & 0.5041 & 0.5712 & 0.5660 & 0.0083 & -0.0055 & 0.1664 & 0.1663 \\
        SGL* & 0.5297 & 0.6054 & 0.4578 & 0.4888 & & 0.5071 & 0.4912 & 0.5558 & 0.5531 & -0.0003 & -0.0223 & 0.1325 & 0.0908 \\
        \cmidrule{1-5} \cmidrule{7-14}
        Return & 0.0389 & 0.0621 & 0.0274 & 0.0368 & &0.3065 & 0.3438 & 0.3403 & 0.3883 & -0.1747 & -0.1819 & -0.5236 & -0.4699 \\
        Sharpe & 0.0453 & 0.0665 & 0.0324 & 0.0411 & &0.4137 & 0.4174 & 0.4743 & 0.4667 & -0.0832 & -0.1011 & -0.1269 & -0.1362 \\
        \cmidrule{1-5} \cmidrule{7-14}
        two-step* & 0.2767 & 0.3834 & 0.2193 & 0.2629 & &0.4479 & 0.4425 & 0.5526 & 0.5743 & -0.0227 & -0.0335 & 0.1457 & 0.1849 \\
        MVECF* & 0.2170 & 0.2321 & 0.2025 & 0.2087 & &0.4286 & 0.4149 & 0.5081 & 0.5068 & -0.0426 & -0.0644 & -0.0281 & -0.0482 \\
        \cmidrule{1-5} \cmidrule{7-14}
        DyRep & 0.3047 & 0.4533 & 0.2243 & 0.2852 & &0.4581 & 0.4499 & 0.5383 & 0.5403 & -0.0235 & -0.034 & 0.0769 & 0.0919 \\
        Jodie & 0.4324 & 0.5757 & 0.3247 & 0.3838 & & 0.5156 & 0.4924 & 0.5757 & 0.5824 & 0.0074 & -0.0022 & 0.2186 & 0.2617 \\
        TGAT & 0.5138 & 0.6318 & 0.4100 & 0.4585 & & \textbf{0.5826} & 0.5423 & \textbf{0.6129} & 0.6037 & \textbf{0.0460} & \underline{0.0343} & \textbf{0.3178} & 0.3452 \\
        TGN & \textbf{0.5673} & \textbf{0.6809} & \underline{0.4611} & \underline{0.5079} & & 0.5405 & 0.5107 & 0.5612 & 0.5506 & 0.0260 & 0.0075 & 0.1959 & 0.1899 \\
        \cmidrule{1-5} \cmidrule{7-14}
        \texttt{PfoTGNRec} & 0.5572 & \underline{0.6674} & 0.4532 & 0.4986 & & \underline{0.5652} & \underline{0.5434} & \underline{0.6125} & \underline{0.6147} & \underline{0.0407} & \textbf{0.0349} & \underline{0.3053} & \textbf{0.3649} \\
        \bottomrule
    \end{tabular}
    }
    {\raggedright \footnotesize \phantom{xx}\textit{Note}: Models with * exclude cold start user results. The best and second best performing models are highlighted in bold and underline, respectively.\par}
    \vspace{0.5em}
    \caption{Comparison of Models Based on Various Metrics}
    \label{tab:model_metrics}
\end{table*}

As shown in Table \ref{tab:model_metrics}, recommender models consistently outperform price-based and stock recommendation models, demonstrating their effectiveness in capturing user preferences. Interestingly, despite the expectation that dynamic models would surpass static models in performance, both types exhibited similar performance. 
To investigate this, we further analyze the results based on testing interactions where items not purchased during the training period were subsequently purchased. This analysis reveals a significant drop in the performance of static recommendation models, while dynamic recommendation models perform markedly better. For example, the best-performing static model, BPR, achieves an NDCG@5 of 0.0416. In contrast, the best-performing dynamic model, TGN, achieves an NDCG@5 of 0.4535. This difference underscores the limitations of static recommendation models, which tend to recommend items that users have already purchased.
In the context of stock recommendation, users may indeed repurchase previously bought items. However, recommending only previously purchased items does not contribute to portfolio diversification. Consequently, dynamic models clearly have an advantage in offering more diverse recommendations.

Compared to dynamic recommender models, our model outperforms other models but falls slightly short of TGN. This is because we intentionally sacrificed a certain level of recommendation performance through mean-variance efficient sampling to enhance the diversification effect. As expected, price-based models and existing stock recommendation models show lower recommendation performance. These models do not effectively capture individual preferences.

\subsection{Portfolio Performance (RQ3)}
In terms of investment performance, the results indicate that our model generally recorded superior performance across most metrics, despite a few exceptions. In particular, the price-based methods, the Return model, and the Sharpe model rank near the bottom in terms of investment performance, demonstrating the difficulty of predicting future prices based on past prices. This highlights the fundamental challenge of stock price prediction. Surprisingly, the Pop model shows high investment performance in some metrics, which appears to be due to the presence of a popularity bias in the data regarding investment performance. However, this model is not suitable as a stock recommendation model because its recommendation performance is very poor. Interestingly, the stock recommendation models, Two-Step and MVECF, have failed to demonstrate competitive investment performance. This is likely due to their inability to effectively manage the dynamic nature of stock features and user behaviors.

\subsection{Hyperparameter Study (RQ4)}

\begin{figure*}
    \centering
    \includegraphics[width=1\linewidth]{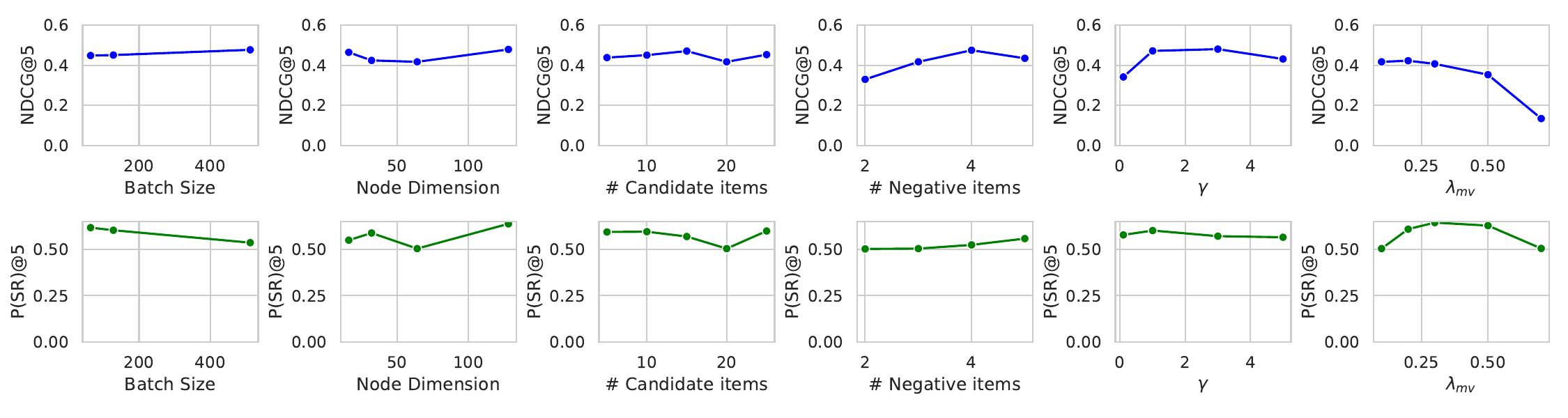}
    \caption{Hyperparameter study of the \texttt{PfoTGNRec}}
    \label{fig:ablation}
\end{figure*}

To thoroughly investigate the impact of various hyperparameters on our model's performance, we conduct an extensive hyperparameter study with six key hyperparameters: batch size, node dimension, number of candidate items, number of negative items, $\gamma$, and $\lambda_{MV}$. For evaluation metrics, we select NDCG@5 to represent recommendation performance and P(SR)@5 to represent portfolio performance. By analyzing these metrics, we aim to derive insights into the trade-offs and interactions between different hyperparameters, thereby guiding the optimization of our model for both recommendation and investment tasks. The results are shown in Figure \ref{fig:ablation}.

\begin{itemize}[left=0pt, label=\textbullet]

\item \textbf{Batch Size}: The analysis reveals that NDCG@5 exhibits a slight increase with larger batch sizes, while P(SR)@5 demonstrates a decreasing trend. This indicates that although larger batches may enhance recommendation quality, they can negatively impact investment performance more than they improve recommendation effectiveness. 

\item \textbf{Node Dimension}: NDCG@5 remains relatively stable across varying node dimensions. However, P(SR)@5 displays notable fluctuations, indicating that the choice of node dimension is critical. Additionally, larger node dimensions can store more information, potentially enhancing performance.

\item \textbf{Number of Candidate Items}: The number of candidate items does not appear to significantly influence NDCG@5. However, P(SR)@5 shows some variability with changes in the number of candidate items. As the number of candidate items increases, the complexity of mean-variance efficient sampling also rises, highlighting the importance of determining an optimal number of candidate items.

\item \textbf{Number of Negative Items}: Both NDCG@5 and P(SR)@5 show an increasing trend with the number of negative items. This positive correlation suggests that incorporating a higher number of negative samples enhances both recommendation and investment performance. However, increasing the number of negative items also may lead to longer computation times, making it crucial to determine an optimal number of negative items.

\item $\boldsymbol{\gamma}$: It exhibits a peak in both NDCG@5 and P(SR)@5 at the value of 1. When performing mean-variance efficient sampling, the weight given to the volatility relative to the return of the stocks changes according to the value of $\gamma$. Therefore, optimizing this hyperparameter is crucial.

\item $\boldsymbol{\lambda_{MV}}$: Our findings reveal that $\lambda_{MV}$ has the most significant impact on both metrics. NDCG@5 decreases sharply with increasing $\lambda_{MV}$, indicating a negative impact on recommendation performance. In contrast, P(SR)@5 shows a peak at a specific $\lambda_{MV}$ value. Although $\lambda_{MV}$ is a hyperparameter that balances recommendation and portfolio performance, the inherent uncertainty in predicting future prices leads to inconsistent impact on investment performance. 

\end{itemize}




\section{Conclusion} 
In this paper, we present \texttt{PfoTGNRec}, a novel framework tailored for stock recommender systems, focusing on two key aspects: capturing the temporal dynamics of the stock market and user preference and integrating portfolio diversification into recommendations. Using temporal graph networks, \texttt{PfoTGNRec} effectively models user preferences that change over time, with a novel training approach specifically designed for portfolio diversification, balancing user preferences with investment risk management.
Experiments have demonstrated \texttt{PfoTGNRec}'s effectiveness, showing competitive recommendation accuracy and improved portfolio performance. 


For future work, we propose incorporating static features of users and items as node features to enhance the model's ability to capture inherent characteristics that influence user behavior and stock performance. Additionally, expanding the recommendation system to account for various user behaviors, such as selling, holding, and ordering will provide a more comprehensive understanding of users and improve the relevance of recommendations. By addressing these areas, we aim to further enhance the robustness and applicability of \texttt{PfoTGNRec} in the real world.

\begin{acks}
This work was supported by National Research Foundation of Korea (NRF) grant (No. NRF-2022R1I1A4069163) and Institute of Information \& Communications Technology Planning \& Evaluation (IITP), South Korea grant (No. RS-2020-II201336, Artificial Intelligence Graduate School Program (UNIST)) funded by the Korea government (MSIT).
\end{acks}

\bibliographystyle{ACM-Reference-Format}
\bibliography{sample}

\end{document}